\documentclass{article}
\usepackage{graphicx}

\usepackage[hidelinks]{hyperref}
\usepackage{url}
\usepackage{xcolor}
\usepackage{natbib}
\usepackage{amsmath,amsthm,amsfonts,amssymb}
\usepackage{tikz}
\usepackage{nicefrac}

\newtheorem{theorem}{Theorem}

\title{Sleeping Kelly}
\author{Ben Abramowitz}
\date{June 2025}

\begin{document}

\maketitle

\begin{abstract}
    The Sleeping Beauty problem is a problem of imperfect recall that has received considerable attention. One approach to resolving the Sleeping Beauty problem has been to allow Sleeping Beauty to make decisions based on her beliefs, and then characterize what it takes for her decisions to be “rational”. In particular, she can be allowed to make monetary bets based on her beliefs, with the assumption that she wants to gain wealth rather than lose it. However, this approach is often coupled with the erroneous assumption that Sleeping Beauty should maximize the expected value of her bets. Here, we show that (1) Sleeping Beauty's \textit{ex ante} optimal betting strategy and \textit{de se} optimal betting strategy are identical, and (2) under an asymptotically optimal betting strategy Sleeping Beauty is invulnerable to diachronic Dutch Books as a Thirder but remains vulnerable as a Halfer.
\end{abstract}

\section{Problem Statement}
The Sleeping Beauty problem was presented as follows by \citet{elga2000self}\footnote{A version of the problem is thought to be first proposed by~\cite{zuboff1990one}}:

\begin{quote}
    ``Some researchers are going to put you to sleep. During the two days that your sleep will last, they will briefly wake you up either once or twice, depending on the toss of a fair coin (Heads: once; Tails: twice). After each waking, they will put you to back to sleep with a drug that makes you forget that waking. When you are first awakened, to what degree ought you believe that the outcome of the coin toss is Heads?"
\end{quote}

\begin{tikzpicture}
  \node[draw, circle] (n1) at (0, 0) {Sleep};
  \node[draw, circle] (n2) at (1.5, 0) {Flip};
  \node[draw, circle] (n3) at (3, 1.5) {Wake};
  \node[draw, circle] (n4) at (3, -1.5) {Wake};
  \node[draw, circle] (n5) at (6, -1.5) {Wake};
  
  \node at (0, 2.5) {\underline{Sunday}};
  \node at (3, 2.5) {\underline{Monday}};
  \node at (6, 2.5) {\underline{Tuesday}};

  \draw[->] (n1) -- (n2);
  \draw[->] (n2) -- (n3) node[midway, above] {H};
  \draw[->] (n2) -- (n4) node[midway, below] {T};
  \draw[->] (n4) -- (n5);
\end{tikzpicture}

\section{Halfers, Thirders, and Others}
The two most common answers to the Sleeping Beauty problem are to assign probabilities of $\frac{1}{2}$ (``Halfer" position) and $\frac{1}{3}$ (``Thirder" position) to the coin having landed heads. \cite{elga2000self} argued for the Thirder position, while \cite{lewis2001sleeping} prominently argued for the Halfer position, but debate continues. According to~\cite{bostrom2007sleeping}, ``neither side seems to have gained a decisive advantage", and \citeauthor{bostrom2007sleeping} argues for a separate answer entirely.

When introducing the problem, \cite{elga2000self} included a basic argument for each position, copied below, before arguing for the Thirder position.

\begin{quote}
    ``First answer: $\frac{1}{2}$, of course! Initially you were certain that the coin was fair, and so initially your credence in the coin's landing Heads was $\frac{1}{2}$. Upon being awakened, you receive no new information (you knew all along that you would be awakened). So your credence in the coin's landing Heads ought to remain $\frac{1}{2}$."
\end{quote}
\begin{quote}
    ``Second answer: $\frac{1}{3}$, of course! Imagine the experiment repeated many times. Then in the long run, about $\frac{1}{3}$ of the wakings would be Heads-wakings -- wakings that happen on trials in which the coin lands Heads. So on any particular waking, you should have credence $\frac{1}{3}$ that that waking is a Heads-waking, and hence have credence $\frac{1}{3}$ in the coin's landing Heads on that trial. This consideration remains in force in the present circumstance, in which the experiment is performed just once."
\end{quote}


Some observers have also argued that the problem is ill-posed or indeterminate. The frequentist argument goes roughly as follows. Suppose you were to repeat the experiment many times, flipping a new coin for each experiment, but each time you wake Sleeping Beauty on Monday or Tuesday you simply ask her whether the coin landed Heads or Tails, rather than asking her to assign a probability. If she always says Heads, then her guess will be correct approximately $\frac{1}{3}$ of the times she is asked, but she will get the correct answer in $\frac{1}{2}$ of the experiments.

\section{Betting Games}

\begin{quote}
    ``It is a platitude among decision theorists that agents should choose their actions so as to maximize expected value. But exactly how to define expected value is contentious."-~\cite{briggs2010decision}
\end{quote}

A class of approaches to the Sleeping Beauty problem allow Sleeping Beauty to make decisions with consequences each time she is woken, rather than simply asking her to state probabilities. In general, she can be made to play a game of some kind, and we can assess her performance in the game~\citep{shaw2013se}. One version of the decision-based approach is to allow Sleeping Beauty to make monetary bets on the coin flip's outcome before she is put to sleep and each time she is woken~\citep{conitzer2015can}.

The trouble with betting arguments is that one must make assumptions about Sleeping Beauty's betting strategy and how it depends on the probabilities she assigns to different events. The existing literature from decision theory on the Sleeping Beauty problem appears to uniformly makes the same assumptions: that it is rational for Sleeping Beauty to either maximize the expected value of each individual bet or maximize the expected value of the sum of her bet payoffs (see for example~\citep{arntzenius2002reflections,halpern2004sleeping, conitzer2015can,conitzer2015devastating,conitzer2015dutch}). Neither assumption is sound.

\section{The PGM Coin Flip}
The core problem with the expectation value assumptions is that they confuse bet payoffs with von Neumann-Morgenstern utilities. 
The difference is revealed lucidly by the infamous coin flip of ~\cite{peters2016evaluating}(PGM).

Suppose you are offered to play a game in which you must wager some amount of money $x$ and then a coin will be flipped. If the coin lands heads, then you will gain 50\% for a payoff of $1.5x$, but if the coin lands tails you will lose 40\% for a payoff of $0.6x$. Naturally, we assume you would strictly prefer winning the bet and increasing your wealth over not betting, and you would strictly prefer not betting to losing the bet and decreasing your wealth. Is it rational for you to play the game?

Suppose we were to say that it is rational for you to maximize the expected value of your payoff. The expected value is $\frac{1.5x + 0.6x}{2} = 1.05x$, or 1.05 times your initial wager. It is commonly thought that this makes it a worthwhile bet because $1.05 > 1$. Since the game is offered in isolation, presumably if it makes sense to play the game, then it makes sense to play the game arbitrarily many times. If you have the opportunity to play the game many, many times, should you?

Suppose you were to play the PGM coin flip game many times in \textit{sequence}. You start with an initial wager of $x$. In each round the coin is flipped your wealth grows or shrinks and becomes your wager in the next round, with no opportunity to add or remove money from the game. For example, suppose in the first two rounds you flip heads twice in a row. Then your wealth would be $x(1.5)^2 = 2.25x$. If the coin flips tails twice in a row, then your wealth would be $x(0.6)^2 = 0.36x$. And finally, if you flip heads once and tails once, in either order, then your wealth would be $x(1.5)(0.6) = 0.9x$ after two rounds. Notice how after an even number of rounds $2n$, if you flip an equal number of heads and tails, then your wealth has decreased to $x(0.9)^n$. You would need to consistently flip more heads than tails just to break even, and you should expect your wealth to decrease at an exponential rate. Playing the sequential game is a bad bet.

Now suppose instead that you were to play the PGM coin flip game many, many times independently in \textit{parallel}. After all of the games conclude, you sum your total winnings (or loses) to get your final payoff. Let $2n$ be the number of games you play, betting $x$ each time. In the limit as $n \rightarrow \infty$, you expect the coin to land heads in approximately half of the games and tails in approximately half of the games. Your expected payoff therefore converges to $x\frac{1.5n + 0.6n}{2n} = 1.05x$. In other words, the expected value of each individual stage game is the ensemble average over many parallel plays of the game in the infinite limit. In this scenario, playing the ensemble of games is a great bet, at least for sufficiently large $n$.

Ultimately, when asked about a single game -- like one bet on one coin flip -- you cannot determine whether or not the bet is worth taking without making further assumptions.

One of the key features of the Sleeping Beauty problem, and other problems of imperfect recall, is that events occur in sequence. To use betting arguments for these problems, we need to be careful in specifying the appropriate wealth dynamics. This is precisely where existing Dutch Book arguments for the Sleeping Beauty problem go wrong. We now examine the Sleeping Beauty problem in which she is allowed to make bets each time she is awakened, recognizing that these actions take place in sequence.

\section{Problem Setup: Sequential Betting}
You are Sleeping Beauty. You begin the experiment with a certain amount of money which you can use to make wagers during the experiment. Some researchers are going to put you to sleep on Sunday. Before you are put to sleep you have the option to bet any fraction of your wealth $[0,1]$ on the outcome of a fair coin flip at 1:1 odds. You do not have access to leverage, so you cannot bet more than 100\% of your wealth. The coin will be flipped after you are put to sleep and you will not be shown the outcome until the experiment is over on Wednesday. If the coin lands heads you will be woken up on Monday. If the coin lands tails you will be woken up on Monday and Tuesday. Each time you are woken up you will have the option to wager any faction of your money on the outcome of the coin flip at 1:1 odds before being put back to sleep. You will not be informed of how much money you have at any point, so all of your wagers must be given as a fraction of your wealth rather than a monetary value. Each time you are put back to sleep your memory will be erased so you have no recollection of being woken or your previous wager if there was one. Your goal is to determine what fraction of your wealth to wager on the coin landing heads or tails before being put to sleep and each time you are woken.

\section{Expected Value versus Wealth Multiplier}
What is the optimal betting strategy in the Sleeping Beauty problem where bets are a fraction of your wealth? Let $-1 \leq a \leq 1$ be the fraction of wealth you wager on the coin landing tails before being put to sleep, and let $-1 \leq b \leq 1$ be the fraction of wealth you wager on the coin landing tails each time you are woken during a run of the experiment. If $a < 0$ or $b < 0$, this means wagering on the coin landing heads instead of tails.

\subsection{Expected Value}
If your goal were to maximize the expected value of your wealth at the end of each experiment, then you would choose $a$ and $b$ to maximize the value $\frac{(1-a)(1-b) + (1+a)(1+b)(1+b)}{2}$. This function is maximized by setting $a=b=1$; going all-in on every wager, to get an expected value of 4 times your original wealth\footnote{This is a polynomial in two variables, so the maxima are found by taking the partial derivatives to get the critical points, and second derivatives to determine the maxima.}\label{footnote_polynomial}. After the experiment is over, you will either have no money left or 8 times your original endowment.

The idea of betting 100\% of your wealth on a coin flip, before being put to sleep or when woken up, should feel wrong. It is.
If you run this experiment many times in sequence, where your wealth at the end of each experiment becomes your wealth at the start of the next, then this strategy guarantees you eventually go bankrupt (with probability 1). Eventually you will lose a bet, and from then on there is no money left to wager.

Similarly, if we consider using this strategy in the ``Extreme Sleeping Beauty" version of the experiment from~\cite{bostrom2007sleeping} where Sleeping Beauty is awakened 1 million days in a row when the coin lands tails rather than only twice, virtual bankruptcy occurs with probability almost  $\frac{1}{2}$ in a single experiment.

\subsection{Wealth Multiplier}\label{sec_wealthmultiplier}
If we plan to repeat the experiment many times in sequence, where the final wealth after one experiment becomes the initial wealth at the next, what is Sleeping Beauty's optimal betting strategy?


Imagine we run the experiment twice in a row, and the coin lands heads once and tails once. Then your wealth at the end of the two experiments is $(1-a)(1-b)(1+a)(1+b)(1+b)$ times your original wealth.\footnotemark[3] To maximize this value we must choose $a=0$ and $b=\frac{1}{3}$.

Of course, if we run the experiment only twice we are not guaranteed to get one heads and one tails from the coin flips. However, as the number of experiments tends to infinitely, we expect half of the flips to be heads and half to be tails across the experiments, by the law of large numbers. The wealth multiplier is therefore $\sqrt{(1-a)(1-b)(1+a)(1+b)(1+b)}$ per experiment, and ($a = 0$, $b=\frac{1}{3}$) is the optimal strategy for long-term wealth maximization. You do not bet any amount before being put to sleep, but each time you are woken it is best to wager $\frac{1}{3}$ of your wealth that the coin landed tails.

\begin{figure}[h!]
    \centering
    \includegraphics[width=0.5\linewidth]{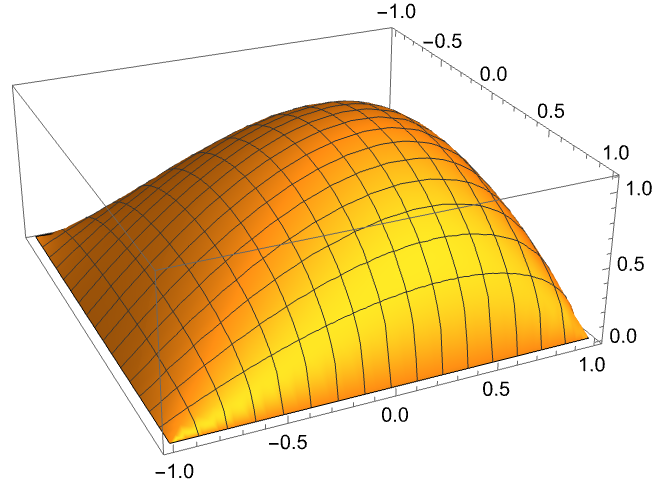}
    \caption{Wealth multiplier for wager fractions $a$ and $b$.}
    \label{fig:growthrate}
\end{figure}

This betting strategy at least passes the smell test. One should not bet any fraction of your wealth on a fair coin toss before the experiment begins. Imagine choosing $a > 0$ and $b=0$, so you are simply betting on successive coin flips. Then by repeating the experiment many times in sequence, your wealth will converge to 0 in the limit for any value of $0 < a \leq 1$, because $(1-a)(1+a) < 1$. Fixing $a=0$, maximizing our wealth multiplier requires us to maximize $(1-b)(1+b)^2$, for which the unique maximum is $b=\frac{1}{3}$.

The asymptotically optimal strategy for sequential bets is referred to as the Kelly criterion, named for its discoverer~\cite{kelly1956new}. We therefore refer to a Sleeping Beauty who maximizes her wealth multiplier as Sleeping Kelly. First, we will show that Sleeping Kelly's betting strategy of $b = \frac{1}{3}$ is invulnerable to Dutch Books. Then we will show that this strategy maximizes the growth rate of Sleeping Beauty's wealth from both the \textit{ex ante} and \textit{de se} perspectives.

\section{Consistency of Betting Strategy}
Recall that we presume Sleeping Kelly rejects all bets before being put to sleep $(a = 0)$ because it is never advantageous to bet any positive amount on a fair coin with fair odds.
Sleeping Kelly's goal is to choose a bet size $b$ for each awakening to maximize her wealth multiplier.

\paragraph{Betting strategy per experiment (\textit{ex ante})}
Looking at experiments, the probability of the coin landing heads in any experiment is $\frac{1}{2}$. The wealth multiplier per experiment is $(1-b)^{\frac{1}{2}} (1+b)^{2 \cdot \frac{1}{2}}$. We can equivalently maximize the logarithm of the wealth multiplier, $\frac{1}{2} \log(1-b) + \frac{1}{2} \cdot 2 \cdot \log (1+b)$. Once again the objective is maximized by $b = \frac{1}{3}$.

\paragraph{Betting strategy per awakening (\textit{de se})}
Looking at awakening events, the wealth multiplier for a single bet upon awakening is $(1-b)^{\frac{1}{3}} (1+b)^{\frac{2}{3}}$, corresponding to the probability of $\frac{1}{3}$ that the coin lands heads. Note that we can equivalently maximize the logarithm of the wealth multiplier, $\frac{1}{3}\log(1-b) + \frac{2}{3}\log (1+b)$. The objective is maximized by $b = \frac{1}{3}$.

\paragraph{Equivalent Optimization Functions}
We can see that whether we reason about our betting strategy upon awakening or before the experiment, after taking the logarithm we ultimately maximize the same function $\log(1-b) + 2 \log (1+b)$ up to rescaling. Whether reasoning from the \textit{ex ante} view assigning a probability of $\nicefrac{1}{2}$ to the coin landing heads, or from the \textit{de se} view assigning a probability of $\nicefrac{1}{3}$ to the coin landing heads upon awakening, Sleeping Kelly ultimately makes the same bets. Her optimal strategy throughout the experiment remains the same as before the experiment began.


\section{Dutch Books}
When deriving $b=\frac{1}{3}$ as the optimal betting strategy, Sleeping Beauty got to decide what fraction of her wealth to wager, and whether to wager it on heads or tails. We now look at a setting in which Sleeping Beauty's strategy space is more restricted. Sleeping Beauty will be given an offer to bet on the outcome of the coin flip before she is put to sleep and each time she is woken, but the experimenter will determine the wager, and Sleeping Beauty only has the option to accept or reject the bet. When put back to sleep, she will have no memory of what previous bets she has been offered or whether she accepted them. We require that the same bet be offered whenever she is woken, and that it be determined before the experiment begins, so the offered bet cannot depend on the coin's outcome or what day it is or Sleeping Beauty's prior decisions. The experimenter can never make choices about what bets to offer using information that is not available to Sleeping Beauty.

The question we want to answer is whether Sleeping Beauty can be subjected to a diachronic Dutch Book. As \cite{conitzer2015dutch} explains,

\begin{quote}
    ``A Dutch book is a set of bets that the agent in question would all accept individually, but that together ensure that the agent incurs a strict loss overall. In a diachronic Dutch book, the bets are offered at different times[.]"
\end{quote}

\subsection{Dutch Book of Expected Value Maximizer}
Different Dutch books have been constructed for Sleeping Beauties who are ``evidential decision theorists" and ``causal decision theorists" that maximize the expected value of their bets~\citep{conitzer2015devastating,conitzer2015dutch}. Consider the following basic example against the Halfer position from~\citet{hitchcock2004beauty}:

Before going to sleep, Sleeping Beauty is offered a wager at almost 1:1 odds that wins $15+\epsilon$ if the coin lands tails and loses $-15$ if the coin lands heads. Each time she is woken, she is offered a bet that wins $10+\epsilon$ if the coin landed heads, and loses $-10$ if the coin landed tails. Both bets have an expected value slightly greater than zero, so Sleeping Beauty is presumed to accept all bets she is offered. However, if the coin lands heads she ends up losing $-5$ and if the coin lands tails she also ends up losing $-5$ by the end of the experiment.

The key flaw here is assuming that a rational Sleeping Beauty would accept any of the individual bets in this example. Observe that if you start with any finite amount of wealth and place a sequence of wagers of the same size one a fair coin with 1:1 payoff odds, then your wealth dynamics follow an unbiased random walk which guarantees eventual bankruptcy. Perturbing the payoff odds by some small $\epsilon$ to give a slightly biased random walk still leaves an overwhelming chance of ruin.

A rational sequential betting strategy for Sleeping Beauty cannot be defined independent of the fraction of her wealth being put at stake. The relation between the size of her bet and current endowment of wealth cannot be ignored when considering her wealth dynamics.
We want to see whether any Dutch book can be created for a Sleeping Beauty who only accepts bets with a wealth multiplier greater than 1.

\subsection{Invulnerability of Sleeping Kelly as a Thirder}
We want to know whether Sleeping Beauty can be subjected to a Dutch book if she is only willing to accept bets that individually have a wealth multiplier greater than 1; i.e., greater than just sitting on the cash and not betting it.

Suppose there is an event with $k$ possible outcomes $\{A_1,\ldots, A_k\}$ with respective probabilities $p_1, \ldots, p_k$. You are offered a bet where your wealth is multiplied by $(1+\alpha_i)$ if event $A_i$ occurs for each $i \leq k$. If you require your wealth multiplier to be greater than one then this means you require $\prod\limits_{i \leq k} (1+\alpha_i)^{p_i} > 1$.

\begin{theorem}
    Sleeping Kelly cannot be subjected to a Dutch book if she is a Thirder and only accepts bets with a wealth multiplier greater than 1.
\end{theorem}

\begin{proof}
Let's define the following variables which all lie in the range $[-1,1]$:

\begin{itemize}
    \item $\alpha_{OH}$ = fraction of wealth won from bet before going to sleep if coin lands heads
    \item $\alpha_{OT}$ = fraction of wealth won from bet before going to sleep if coin lands tails
    \item $\alpha_{WH}$ = fraction of wealth won from bet when woken up if coin lands heads
    \item $\alpha_{WT}$ = fraction of wealth won from bet when woken up if coin lands tails
\end{itemize}

Recall that as a Thirder Sleeping Kelly assigns a probability of $\frac{1}{2}$ to the coin landing tails before being put to sleep and probability $\frac{2}{3}$ to the probability of tails upon being woken.
We now have two conditions for whether Sleeping Beauty accepts a bet. Before sleeping she requires $(1+\alpha_{OH})^{1/2}(1+\alpha_{OT})^{1/2} > 1$ and upon waking  she requires $(1+\alpha_{WH})^{1/3}(1+\alpha_{OT})^{2/3} > 1$. We can simplify the exponents to get our two conditions for bet acceptance.

\vspace{5mm}
\textbf{Bet Acceptance Conditions:}
\begin{itemize}
    \item Before Sleeping: $(1+\alpha_{OH})(1+\alpha_{OT}) > 1$
    \item Upon Waking: $(1+\alpha_{WH})(1+\alpha_{WT})^2 > 1$
\end{itemize}

\vspace{5mm}
To construct a Dutch book, in which Sleeping Beauty loses money regardless of the outcome of the coin flip, we would need a set of values such that the following two conditions hold.

\vspace{5mm}
\textbf{Dutch Book Conditions:}
\begin{itemize}
    \item Losing When Heads: $(1+\alpha_{OH})(1+\alpha_{WH}) < 1$
    \item Losing When Tails: $(1+\alpha_{OT})(1+\alpha_{WT})^2 < 1$
\end{itemize}

\vspace{5mm}
If both of the Dutch book conditions hold, then by multiplying them together, we see that the wealth multiplier must be less than 1: $$(1+\alpha_{OH})(1+\alpha_{WH})(1+\alpha_{OT})(1+\alpha_{WT})^2 < 1.$$ However, this is not possible  according to our bet acceptance conditions. By reordering the terms, we see $$\underbrace{(1+\alpha_{OH})(1+\alpha_{OT})}_{>1}\underbrace{(1+\alpha_{WH})(1+\alpha_{WT})^2}_{>1} > 1$$

Thus, Sleeping Kelly cannot be subjected to Dutch books if she is Thirder.
\end{proof}

\section{Vulnerability of Halfers}
If Sleeping Kelly approaches each individual bet upon awakening as a Halfer, assigning a probability of heads $\frac{1}{2}$, then she becomes vulnerable to Dutch books.

\begin{theorem}
    Sleeping Kelly is vulnerable to Dutch books if she assigns a probability of $\frac{1}{2}$ to the coin landing heads or tails each time she is woken, even if she only accepts individual bets with a wealth multiplier greater than 1.
\end{theorem}

Now we have a different set of bet acceptance conditions, but the same Dutch book conditions as before.

\vspace{5mm}
\textbf{Bet Acceptance Conditions:}
\begin{center}
\begin{itemize}
    \item Before Sleeping: $(1+\alpha_{OH})(1+\alpha_{OT}) > 1$
    \item Upon Waking: $(1+\alpha_{WH})(1+\alpha_{WT}) > 1$
\end{itemize}
\end{center}

\vspace{5mm}
\textbf{Dutch Book Conditions:}
\begin{itemize}
    \item Losing When Heads: $(1+\alpha_{OH})(1+\alpha_{WH}) < 1$
    \item Losing When Tails: $(1+\alpha_{OT})(1+\alpha_{WT})^2 < 1$
\end{itemize}

\vspace{5mm}
Now a Dutch book is possible. For example we can offer the Halfer the following bets for any suitably small $\epsilon$: 

\begin{itemize}
    \item $\alpha_{WH} = \frac{1}{2}$
    \item $\alpha_{WT} = -\frac{1}{3}+\epsilon$
    \item $\alpha_{OT} = 1$
    \item $\alpha_{OH} = -\frac{1}{3}-\epsilon$
\end{itemize}

Before sleeping, her wealth multiplier would be $\frac{4}{3}-\epsilon$ and upon waking her wealth multiplier would be $1+\epsilon$. However, if the coin lands heads then her wealth shrinks by a factor of $1 - \epsilon$, and if the coin lands tails then her wealth shrinks by a factor of $\frac{8}{9}-\epsilon$.

\section{Conclusion}
If Sleeping Beauty is allowed to place bets on the outcome of the coin flip before being put to sleep and each time she is awakened, then by maximizing her wealth multiplier, her betting strategy remains consistent between the \textit{ex ante} and \textit{de se} views. By maximizing the growth rate of her wealth according to the Kelly criterion, Sleeping Beauty's remains invulnerable to Dutch books. This betting argument stands in contrast to the assumption that a bettor rationally maximizes the expected payoff of their bets per experiment or per awakening.

\bibliography{bib}
\bibliographystyle{plainnat}

\end{document}